\documentclass[12pt]{article}

\usepackage{jheppub}
\usepackage{amsbsy,amsfonts,amsmath,amssymb}
\usepackage{cleveref} 
\usepackage[g]{esvect} 

\crefname{equation}{}{}
\crefname{chapter}{Chapter}{Chapters}
\crefname{section}{Section}{Sections}
\crefname{subsection}{Subsection}{Subsections}
\crefname{subsubsection}{Subsubsection}{Subsubsections}
\crefname{figure}{Figure}{Figures}
\crefname{table}{Table}{Tables}
\crefname{appendix}{Appendix}{Appendices}

\allowdisplaybreaks

\begin{document}

\title{The dynamical second-order transport coefficients of smeared Dp-brane}

\author[]{Chao Wu, Yanqi Wang and Yi-An Wang}

\affiliation[]{School of Physics and Optoelectronic Engineering, Anhui University, Hefei 230601, China}

\emailAdd{chaowu@ahu.edu.cn, wangyanqi0@gmail.com, yianwang777@outlook.com}

\abstract{The smeared Dp-brane is constructed by having the black Dp-brane uniformly smeared over several transverse directions. After integrating the spherical directions and the smeared directions, the smeared Dp-brane turns out to be a Chamblin-Reall model with one background scalar field. Within the framework of the fluid/gravity correspondence, we not only prove the equivalence between the smeared Dp-brane and the compactified Dp-brane by explicitly calculating the 7 dynamical second-order transport coefficients of their dual relativistic fluids, but also revisit the Correlated Stability Conjecture for the smeared Dp-brane via the fluid/gravity correspondence.}



\maketitle

\section{Introduction}

Holographic explorations on the transport properties of non-conformal relativistic fluids have kept growing in the last dozen years \cite{Gubser0804PRL,Gubser0804PRD,Gubser0806,Finazzo1412,Attems1603,Kleinert1610,Li1411,
Buchel0311,Buchel0406200,Benincasa0507,Buchel0812,Buchel0908,Buchel1110,Buchel1503,Bigazzi0909,
Bigazzi0912,Buchel0509,Buchel0903,Bigazzi1006,Parnachev0506,Benincasa0605,Mas0703,Natsuume0807,
Springer0810,Springer0902,Kanitscheider0901,David0901,Wu1508,Wu1604,Wu1807,Wu2012,Wu2111,
Wu1608}. A recent article \cite{Wu2111} makes an interesting observation that all the classical backgrounds of the gravity side in the above-mentioned studies leading to non-conformal and analytical results
\cite{Buchel0311,Buchel0406200,Benincasa0507,Buchel0812,Buchel0908,Buchel1110,Buchel1503,
Bigazzi0909,Bigazzi0912,
Buchel0509,Buchel0903,
Bigazzi1006,Kleinert1610,Parnachev0506,Benincasa0605,Mas0703,Natsuume0807,
Springer0810,Springer0902,Kanitscheider0901,David0901,Wu1508,Wu1604,Wu1807,Wu2012,Wu2111,
Wu1608} are the Chamblin-Reall type \cite{Chamblin9903}. To make one further step, it classifies the  Chamblin-Reall backgrounds by the number of independent scalar fields, with the number running from 1 to 4. Only the backgrounds with one scalar can be solved both exactly and analytically. These backgrounds are the NS5- \cite{Parnachev0506},
Dp- \cite{Mas0703,Natsuume0807,Springer0810,Springer0902,Kanitscheider0901,David0901}
and compactified Dp-branes \cite{Benincasa0605,Wu1508,Wu1604,Wu1807,Wu2012,Wu1608}\footnote{The model in ref. \cite{Wu1608} is a little special in that the background also has a vector field. It can also be solved both exactly and analytically, despite the independent scalar fields are three.}
from type II superstring theory, as well as the reduced compactified AdS black hole \cite{Bigazzi1006,Kleinert1610,Wu2111}\footnote{Ref. \cite{Kleinert1610} is mainly on the numerical study of the holographic renormalization flow model \cite{Attems1603}. Whereas it contains the exact and analytical results of the 5-dimensional reduced AdS black hole in the appendix.}.

Here we find another 10-dimensional background from superstring theory which, after dimensional reduction, also leads to a Chamblin-Reall background with one scalar field. It is the smeared Dp-brane that is constructed by having the Dp-brane smeared on several of its transverse directions. After it was proposed in ref. \cite{Harmark0204}, the smeared Dp-brane was used to discuss the relation between dynamical and thermodynamical instabilities \cite{Bostock0405}. The story about the dynamical stability of the gravitational systems begins with refs. \cite{Gregory9301,Gregory9404}, which discovered some Dp-branes in superstring theory are unstable under classical metric perturbations. Then after checking the charged AdS black holes, refs. \cite{Gubser0009,Gubser0011} found  thermodynamical instabilities usually indicate the presence of dynamical instabilities, which is called the Gubser-Mitra conjecture or the Correlated Stability Conjecture (CSC) \cite{Harmark0509} later on. However, a study on the metric perturbation in the smeared direction shows the uniformly smeared Dp-brane does not obey the CSC \cite{Bostock0405}. But the subsequent investigations \cite{Aharony0406,Gubser0411,Ross0503} point out that this inconsistency is caused by using the wrong ensemble. One should use the grand canonical ensemble but not canonical since the brane charge can vary with respect to the metric perturbation. Soon after further explorations on the D-brane bound states \cite{Friess0503} and the Maldacena-Nunez background \cite{Buchel0507} that supports the CSC, counter examples are found in ref. \cite{Friess0508} which stop the research on the CSC for a couple of years.

With the advent of the world-volume effective theory of higher dimensional black holes \cite{Emparan0902},  the CSC can then be tested in the framework of blackfold \cite{Emparan0910}. In ref. \cite{Emparan1205}, the authors point out that the previous problem about the CSC is caused by a wrong understanding of the nature of the dynamical instability. The stability of the horizon is of ghost type, not of tachyon type, and the ghost type instability belongs to the hydrodynamical modes inside the translationally invariant horizon. So it should be the hydrodynamical stability that is correlated with the local thermodynamical stability, rather than the metric perturbative stability.

The fluid/gravity correspondence is a fantastic framework to study the thermal and hydrodynamical properties of gravitational background with infinite planar horizon. Our motivation is to revisit the stability problem of the smeared Dp-brane with the final correct understanding of the CSC \cite{Emparan1205} and the new method of fluid/gravity correspondence. We will denote the Dp-brane with $q$ smeared directions the D(p+q)-brane. The cases that we consider in this paper are the D3-brane delocalized on 1 transverse direction; D2-brane with 1 or 2 smeared directions and D1-brane with 1, 2 or 3 smeared directions. The D(p-q)-brane will still be used for Dp-brane with $q$ world-volume dimensions compactified, as in \cite{Wu2012}.

With this motivation, we will calculate the dynamical second-order transport coefficients of smeared Dp-brane. Through the results, the equivalence between the smeared Dp-brane and the compactified Dp-brane will be clearly seen, and the correlation between thermal and hydrodynamical stabilities can also be easily shown. We will also study the smeared D0-brane and look at its thermal and hydrodynamical instabilities from our new viewpoint.

\section{The smeared Dp-brane as Chamblin-Reall background}

The smeared Dp-brane is a kind of 10-dimensional supergravity background of type II string theory, that the Dp-brane delocalizes in one or several of its transverse directions. The delocalization can be either uniform or non-uniform. We only consider the uniformly smeared Dp-brane in this paper.

The 10-dimensional classical action of the Dp-brane which is uniformly smeared in $q$ transverse directions can be written as
\begin{align}\label{eq: smeared Dp 10D action}
  S =&\; \frac1{2\kappa_{10}^2} \int d^{10}x \sqrt{-G} \left[ \mathcal R - \frac12 (\nabla_{\hat M}\phi)^2 - \frac{g_s^2}{2(8-p)!} e^{\frac{p-3}2\phi} \tilde F_{\hat M_1\cdots \hat M_{8-p}}^2 \right] \cr
  & - \frac{1}{\kappa_{10}^2} \int d^9x \sqrt{-H} \mathcal K + \frac{1}{\kappa_{10}^2} \int d^9x \sqrt{-H} \frac{9-p-q}{2L_p} e^{-\frac{(p-3)^2 + q(p+1)}{4(p-3)(7-p-q)}\phi}.
\end{align}
The three terms on the right-hand-side of the equal sign are separately the bulk term, the Gibbons-Hawking surface term, and the counter term. Here we use the magnetic component of the Ramond-Ramond (RR) field $\tilde F_{8-p}$ in the bulk term, the electric component of the RR field will bring some problems in the reducing procedure of the 10-dimensional action. We construct the counter term ourselves to eliminate the divergence in the boundary stress tensor.

The background that solves \cref{eq: smeared Dp 10D action} reads:
\begin{align}
   ds^2 &= \left[ \left( \frac r{L_p} \right)^\frac{(7-p)(7-p-q)}8 \Big( -f(r)dt^2 + \delta_{ij}dx^idx^j \Big) + \left( \frac r{L_p} \right)^{- \frac{(p+1)(7-p-q)}8} \frac{dr^2}{f(r)} \right]  \cr
  & + \left( \frac r{L_p} \right)^{- \frac{(p+1)(7-p-q)}8} \Big( \delta_{mn}dy^mdy^n +  r^2 d\Omega_{8-p-q}^2 \Big),
  \label{eq: smeared Dp 10D metric} \\
   e^\phi &= \left( \frac r{L_p} \right)^\frac{(p-3)(7-p-q)}4, \qquad\qquad \tilde F_{y^1 \cdots y^q \theta^1 \cdots \theta^{8-p-q}} = g_s^{-1} Q_p\sqrt{\gamma_{8-p-q}}, \label{eq: smeared Dp RR field}
\end{align}
where
\begin{align}
  f(r) = 1 - \frac{r_H^{7-p-q}}{r^{7-p-q}}, \quad Q_p = (7-p-q) L_p^{7-p-q}, \quad L_p^{7-p-q} = \frac{(2\pi l_s)^{7-p-q} g_s N}{(7-p-q) \Omega_{8-p-q}}.
\end{align}
Here $g_s$, $l_s$, and $N$ are separately the string coupling, the string length scale, and the number of branes. $\Omega_{8-p-q}$ is the volume of the $(8-p-q)$-dimensional unit sphere. We have split the metric into 3 parts: the world-volume directions of the brane with coordinate $x^M = \{x^\mu, r\} = \{t, x^i, r\}$, the smeared directions parameterized by $y^m$ and the transverse spherical part denoted by $\theta^a$. Then the coordinate for the whole metric is $x^{\hat M} = \{x^M, y^m, \theta^a\}$. The metric, dilaton, and the RR field are all written in the near horizon limit.

In the original construction, the smeared Dp-brane only has one smeared direction \cite{Harmark0204}, but we have set $q$ smeared directions in the metric \cref{eq: smeared Dp 10D metric}, in order to make connections to the compactified Dp-brane \cite{Wu1604,Wu2012}. The difference in \cref{eq: smeared Dp 10D metric} from that of the compactified Dp-brane is that the $y^m$ part is now the transverse directions of the brane in which the D-branes are smeared. If we compactify the smeared directions, then \cref{eq: smeared Dp 10D metric} has topological structure as $\mathcal M_{p+2} \times \mathbf T^q \times \mathbf S^{8-p-q}$.

We use the bulk ansatz
\begin{align}
  ds^2 &= e^{2\alpha_1 A} g_{MN} dx^M dx^N + e^{2\alpha_2 A}\left( e^{2\beta_1 B} \delta_{mn}dy^m dy^n + e^{2\beta_2 B} L_p^2 d\Omega_{8-p-q}^2 \right)
\end{align}
and its boundary version
\begin{align}
  ds^2 &= e^{2\alpha_1 A} h_{MN} dx^M dx^N + e^{2\alpha_2 A}\left( e^{2\beta_1 B} \delta_{mn}dy^m dy^n + e^{2\beta_2 B} L_p^2 d\Omega_{8-p-q}^2 \right)
\end{align}
to reduce the action into $(p+2)$-dimensional. The parameters in the reduction ansatz are chosen to be
\begin{align}
  \alpha_1 = \frac{p-8}{p}, \qquad \alpha_2 = 1, \qquad \beta_1 = - \frac{8-p-q}{q}, \qquad \beta_2 = 1.
\end{align}
To perform the dimensional reduction, we just need to integrate the smeared directions $y^m$ and the spherical directions $\theta^a$, then we have the $(p+2)$-dimensional reduced action
\begin{align}\label{eq: p+2 reduced action}
  S =&\; \frac1{2\kappa_{p+2}^2} \int d^{p+2}x \sqrt{-g} \left[ R - \frac12 (\partial \phi)^2 - \frac{8(8-p)}{p} (\partial A)^2 \right.  \cr
  &\left. - \frac{(8-p)(8-p-q)}{q} (\partial B)^2 - V(\phi,A,B) \right] - \frac1{\kappa_{p+2}^2} \int d^{p+1}x \sqrt{-h} K \cr
  & + \frac1{\kappa_{p+2}^2} \int d^{p+1}x \sqrt{-h} \frac{9-p-q}{2L_p} \exp\left[ - \frac{8-p}{p} A - \frac{(p-3)^2+q(p+1)}{4(p-3)(7-p-q)} \phi \right], \cr
  V(\phi,A,B) =&\; \frac{(7-p-q)^2}{2L_p^2} \exp\left[ \frac{p-3}{2} \phi - \frac{2(p+1)(8-p)}{p}A\right] \cr
   & - \frac{(7-p-q)(8-p-q)}{L_p^2} \exp\left( - \frac{16}{p}A - 2B \right),
\end{align}
with
\begin{align}
\frac{1}{2\kappa_{p+2}^2} = &\; \frac{L_p^{8-p-q}\Omega_{8-p-q}V_q}{2\kappa_{10}^2}.
\end{align}
In the scalar potential of \cref{eq: p+2 reduced action}, we do not have the RR field $\tilde F_{8-p}$, since it is in the $y^m$ and $\theta^a$ directions and has been integrated.

The $(p+2)$-dimensional reduced background has 3 scalar fields and is written as
\begin{align}
  ds^2 &= \left( \frac r{L_p} \right)^{\frac{9-p-q}{p}} \left( -f(r) dt^2 + d \vv x^2 \right) + \left( \frac{r}{L_p} \right)^\frac{(p^2-8p+9) + q(p-1)}{p} \frac{dr^2}{f(r)},\\
  e^{\phi} &= \left( \frac r{L_p} \right)^\frac{(p-3)(7-p-q)}{4}, \\
  e^{A} &= \left( \frac r{L_p} \right)^{\frac{(p-3)^2}{16} - \frac{q (p^2-7p+8)}{16 (8-p)}}, \qquad \qquad e^B = \left( \frac r{L_p} \right)^{\frac{q}{8-p}}.
\end{align}
The 3 scalar fields are not independent since we have
\begin{align}
  A &= \frac{4}{(p-3) (7-p-q)}\left[ \frac{(p-3)^2}{16} - \frac{q (p^2-7p+8)}{16 (8-p)} \right] \phi, \cr
  B &= \frac{4q}{(p-3) (8-p) (7-p-q)} \phi.
\end{align}
Using the above relations, we can rewrite the action as
\begin{align}
  S =&\; \frac1{2\kappa_{p+2}^2} \int d^{p+2}x \sqrt{-g} \left[ R - \frac{4(9-p-q)((p-3)^2 + q(p-1))}{p(p-3)^2 (7-p-q)^2} (\partial \phi)^2 \right. \cr
   &\left. + \frac{(7-p-q) (9-p-q)}{2L_p^2} \exp \left( -\frac{4(p-3)^2 + 4q(p-1)}{p(p-3) (7-p-q)} \phi \right) \right]  \cr
   & - \frac1{\kappa_{p+2}^2} \int d^{p+1}x \sqrt{-h} \left[ K - \frac{9-p-q}{2L_p} \exp \left( - \frac{2(p-3)^2 + 2q (p-1)}{p (p-3) (7-p-q)} \phi \right) \right]. \quad
\end{align}

In order to recast the reduced theory into Chamblin-Reall form, we rescale the scalar field as
\begin{align}
  \sqrt{\frac{4(9-p-q) ( (p-3)^2 + q(p-1) )}{p (p-3)^2 (7-p-q)^2}} \phi = \frac{1}{\sqrt2} \varphi.
\end{align}
Then one can rewrite the reduced theory in the Chamblin-Reall form
\begin{align}
  S =&\; \frac1{2\kappa_{p+2}^2} \int d^{p+2}x \sqrt{-g} \left[ R - \frac12 (\partial \varphi)^2 + \frac{(7-p-q)(9-p-q)}{2L_p^2} e^{-\gamma \varphi} \right] \cr
   & - \frac1{\kappa_{p+2}^2} \int d^{p+1}x \sqrt{-h} \left( K - \frac{9-p-q}{2L_p} e^{ - \frac{\gamma }{2} \varphi } \right).
\end{align}
Then the equations of motions (EOM) can be derived as
\begin{align}
  &E_{MN}-T_{MN}=0,  \cr
  &\nabla^2\varphi - \frac{\gamma}{2L_p^2} (7-p-q) (9-p-q) e^{-\gamma\varphi} = 0
\end{align}
where
\begin{align}
  T_{MN} = \frac12 \left(\partial_M \varphi \partial_N \varphi - \frac12 g_{MN} (\partial \varphi)^2 \right) + \frac1{4L_p^2} (7-p-q) (9-p-q) g_{MN} e^{-\gamma \varphi}
\end{align}
is the energy-momentum tensor in the bulk. The above EOM will be solved by the background written in Chamblin-Reall form
\begin{align}
  ds^2 &= \left( \frac r{L_p} \right)^{\frac{9-p-q}{p}} \left( -f(r) dt^2 + d \vv x^2 \right) + \left( \frac{r}{L_p} \right)^\frac{(p^2-8p+9) + q(p-1)}{p} \frac{dr^2}{f(r)}, \\
  e^{\varphi} & = \left( \frac r{L_p} \right)^{\frac{9-p-q}{2} \gamma}.
\end{align}
Here $\gamma^2 = \frac{2(p-3)^2+2q(p-1)}{p(9-p-q)}$. Based on the discussion in \cite{Wu2111}, we know that this will be the numerical part of the bulk viscosity which will be verified by the end of the first order calculation. The Hawking temperature can be deduced from the above metric directly as
\begin{align}
  T = \frac{7-p-q}{4\pi} \frac{r_H^{\frac{5-p-q}{2}}}{L_p^\frac{7-p-q}{2}}.
\end{align}
We will set $L_p=1$ from now on.

\section{The first order results}

We set the global on-shell metric as
\begin{align}\label{eq: global on-shell metric}
  ds^2 =& -r^\frac{9-p-q}{p} [ f(r_H(x),r) + k(r_H(x), u^\alpha(x), r) ] u_\mu(x) u_\nu(x) dx^\mu dx^\nu \cr
  &- 2 r^\frac{9-p-q}{p} P_\mu^\rho(u^\alpha(x)) w_\rho(u^\alpha(x), r) u_\nu(x) dx^\mu dx^\nu  \cr
  & + r^\frac{9-p-q}{p} [ P_{\mu\nu}(u^\alpha(x)) + \alpha _{\mu\nu}(r_H(x), u^\alpha(x), r) + h(r_H(x), u^\alpha(x), r) P_{\mu\nu}(u^\sigma(x)) ] dx^\mu dx^\nu \cr
  & - 2 r^\frac{(p-3)(p-6) + q(p-2)}{2p} [ 1 + j(r_H(x), u^\alpha(x), r) ] u_\mu(x) dx^\mu dr
\end{align}
where $x^0 = v$ is the Eddington-Finkelstein coordinate and it is related to $t$ by $dt = dv - \frac{dr}{r^\frac{7-p-q}{2} f(r)}$.

The first order expanded on-shell metric can be got by expanding the global on-shell metric to first order:
\begin{align}
  ds^2 =&\; r^\frac{9-p-q}{p} \bigg[ - \bigg( f(r) - \frac{(7-p-q)r_H^{6-p-q}}{r^{7-p-q}}\delta r_H + k^{(1)}(r) \bigg) dv^2 \cr
  & + 2 \big( (f-1) \delta \beta_i + w^{(1)}_i(r) \big) dvdx^i  + (\delta_{ij} + \alpha_{ij}^{(1)}(r) + h^{(1)}(r) \delta_{ij}) dx^idx^j \bigg] \cr
  & + 2 r^\frac{(p-3)(p-6) + q(p-2)}{2p} (1 + j^{(1)}(r)) dvdr - 2 r^\frac{(p-3)(p-6) + q(p-2)}{2p} \delta\beta_i dx^idr
\end{align}
In the above, $\delta r_H = x^\mu \partial_\mu r_H$ and $\delta\beta_i = x^\mu \partial_\mu \beta_i$ as they are in previous works \cite{Wu1604,Wu1608,Wu1807,Wu2012,Wu2111}.

The traceless symmetric tensor part of Einstein equation
\begin{align}
  E_{ij} - \frac1{p} \delta_{ij} \delta^{kl} E_{kl} - \left( T_{ij} - \frac1{p} \delta_{ij} \delta^{kl} T_{kl} \right) &= 0
\end{align}
gives the differential equation for $\alpha^{(1)}_{ij}$ as
\begin{align}
  \partial_r (r^{8-p-q} f(r) \partial_r \alpha^{(1)}_{ij}(r)) + (9-p-q) r^\frac{7-p-q}2 \sigma_{ij} = 0.
\end{align}
In our previous works such as \cite{Wu1807,Wu2012,Wu2111}, the first-order tensor perturbations are solved case by case in specified $p$ and $q$. Here we find that the solution can be expressed via hypergeometric function in general $p$ and $q$, that is $\alpha^{(1)}_{ij} = F(r) \sigma_{ij}$ where
\begin{align}
  F(r) =&\; \frac{4}{(23-3p-3q) r^\frac{5-p-q}{2}} {}_2F_1\left( 1, \frac32+\frac{1}{7-p-q}, \frac52+\frac{1}{7-p-q}, \left( \frac{r_H}{r} \right)^{7-p-q} \right) \cr
  &+ \frac{2 \ln f(r)}{(7-p-q) r_H^\frac{5-p-q}{2}}.
\end{align}
Here ${}_2F_1\left( a,b;c,x \right) = \sum_{n=0}^{\infty} \frac{(a)_n (b)_n}{(c)_n} \frac{x^n}{n!}$ is the hypergeometric series. As one can easily check that $F(r)$ goes to 0 as $r\to \infty$, which also suggests that $p+q<5$ must hold. This means that there are only 6 physically reasonable cases of the smeared Dp-brane that are dual to relativistic fluids, which can be listed as: For $p=1$, $q$ can take 1, 2 or 3; for $p=2$, $q=$ 1, 2; when $p=3$, $q$ can only be 1. $F(r)$ is also regular at $r=r_H$, the detail of the proof can be found in the appendix.

The first-order dynamical equation in the vector part is derived from $E_{ri}-T_{ri} = 0$, which reads
\begin{align}
  \partial_r \left( r^{8-p-q} \partial_r w^{(1)}_i \right) + \frac{9-p-q}{2} r^\frac{7-p-q}{2} \partial_0 \beta_i = 0.
\end{align}
The solution of the above is
\begin{align}
   w_i^{(1)}(r) = a(r) \partial _0\beta_i, \qquad a(r) = \frac{2}{(5-p-q) r^\frac{5-p-q}2}.
\end{align}
The first-order vector constraint equation is derived from $g^{r0} (E_{0i}-T_{0i}) + g^{rr} (E_{ri}-T_{ri}) = 0$, from which one gets
\begin{align}
   \frac{1}{r_H} \partial _i r_H = - \frac{2}{5-p-q} \partial _0 \beta_i.
\end{align}

The first scalar constraint is got from $g^{rr} (E_{rr} - T_{rr}) + g^{r0} (E_{r0} - T_{r0}) = 0$ as
\begin{align}
  \frac1{r_H} \partial_0 r_H = - \frac2{9-p-q} \partial \beta.
\end{align}
The first-order scalar perturbations are solved from $g^{rr} (E_{rr} - T_{rr}) + g^{r0} (E_{r0} - T_{r0}) = 0$, $E_{rr}-T_{rr} = 0$ and the EOM of $\varphi$. They separately give the differential equations of the scalar perturbations as
\begin{align}
   & (r^{7-p-q} k_{(1)})' - 2 (7-p-q) r^{6-p-q} j_{(1)} + \left[ p r^{7-p-q} - \frac{2p}{9-p-q} r_H^{7-p-q} \right] h'_{(1)} \cr
   & + 2 r^\frac{7-p-q}{2} \partial \beta = 0,  \\
   & r h''_{(1)} + \frac{7-p-q}{2} h'_{(1)} - \frac{9-p-q}{p} j'_{(1)} = 0,  \\
   & (r^{7-p-q} k_{(1)})' - r^{7-p-q} f j'_{(1)} - 2(7-p-q) r^{6-p-q} j_{(1)} + \frac{p}{2} r^{7-p-q} f h'_{(1)} \cr
   & + r^\frac{7-p-q}2 \partial \beta = 0.
\end{align}
The solutions are
\begin{align}
  F_h &= \frac1{p} F, \qquad F_j = - \frac{2}{9-p-q} \frac{ r^\frac{9-p-q}{2} - r_H^\frac{9-p-q}{2} }{ r^{7-p-q} - r_H^{7-p-q} } + \frac{5-p-q}{2(9-p-q)} F, \cr
  F_k &= - \frac4{ (9-p-q) r^\frac{5-p-q}{2} } + \frac{1}{9-p-q} \left( 5-p-q + \frac{2r_H^{7-p-q}}{r^{7-p-q}} \right) F.
\end{align}

From the solutions of the first-order differential equations, we can find the $q$-dependence of the smeared brane is quite different from the (compactified) Dp-brane and the compactified AdS black hole. In the smeared brane case, $F_h$ does not depend on $q$ while others all depend on $q$. But in the (compactified) Dp-brane and the compactified AdS black hole, only $F_h$ is $q$-dependent while the others are not. What's more, if one sets $p \to p-q$ in the differential equations of the smeared brane, one will get the corresponding results of the compactified Dp-brane. On the contrary, if setting $p \to p+q$ in the compactified Dp-brane, one has the results of the smeared Dp-brane. This may suggest that the smeared Dp-brane and the compactified Dp-brane are related to each other. We will see this more clearly from the results of transport coefficients.

The Brown-York tensor are defined as
\begin{small}
\begin{align}
  T_{\mu\nu} = \frac{1}{\kappa_{p+2}^2} \lim_{r\to\infty} \left( \frac r{L_p} \right)^\frac{(9-p-q)(p-1)}{2p} \left[ K_{\mu\nu} - h_{\mu\nu}K - \frac{9-p-q}{2{L_p}} \left( \frac r{L_p} \right)^{-\frac{(p-3)^2 + q(p-1)}{2p}} h_{\mu\nu} \right],
\end{align}
\end{small}
from which one can derive the boundary stress-energy tensor of the first order as
\begin{align}
  T_{\mu\nu} =&\, \frac{1}{2 \kappa_{p+2}^2} \left[ {r_H^{7-p} \over L_p^{8-p}} \left( \frac{9-p-q}{2} u_\mu u_\nu + \frac{5-p-q}{2} P_{\mu\nu} \right) \right. \cr
  &\left. - \left( \frac{r_H}{L_p} \right) ^\frac{9-p-q}{2} \left( 2\sigma_{\mu\nu} + \frac{2(p-3)^2 + 2q(p-1)}{p(9-p-q)} P_{\mu\nu} \partial u \right) \right].
\end{align}
Then the thermal quantities and the first-order transport coefficients can be read as
\begin{align}
  \varepsilon &= \frac{1}{2 \kappa_{p+2}^2} \frac{9-p-q}{2} {r_H^{7-p-q} \over L_p^{8-p-q}}, \qquad \mathfrak p = \frac{1}{2 \kappa_{p+2}^2} \frac{5-p-q}{2} {r_H^{7-p-q} \over L_p^{8-p-q}} \cr
  \eta &= \frac{1}{2 \kappa_{p+2}^2} \left( \frac{r_H}{L_p} \right)^\frac{9-p-q}{2}, \qquad \zeta = \frac{1}{2 \kappa_{p+2}^2} \frac{2(p-3)^2 + 2q(p-1)}{p(9-p-q)} \left( \frac{r_H}{L_p} \right)^\frac{9-p-q}{2}.
\end{align}
Then we can calculate the entropy density, the sound speed and the heat capacity as
\begin{align}\label{eq: s c_s and c_V}
  s =&\; \frac{\varepsilon+\mathfrak p}{T} = \frac{1}{2\kappa_{p+2}^2} 4 \pi \left( \frac{r_H}{L_p} \right) ^ \frac{9-p-q}{2}, \qquad c_s^2 = \frac{d \mathfrak p}{d \varepsilon} = \frac{5-p-q}{9-p-q}, \cr
  c_V =&\; \frac{d \varepsilon}{dT} = \frac{1}{2\kappa_{p+2}^2} \frac{4 \pi (9-p-q)}{5-p-q} \left( \frac{r_H}{L_p} \right)^ \frac{9-p-q}{2}.
\end{align}
Previous studies that discuss the CSC about smeared Dp-brane \cite{Ross0503,Bostock0405,Harmark0509} all relate the local thermodynamical stability with the classical stability of metric perturbations, which is not correct by ref. \cite{Emparan1205}.
It suggests that the local thermal stability should be related to hydrodynamical stability. This can be seen via the relation \cite{Emparan1205}
\begin{align}
  c_s^2 = \frac{s}{c_V},
\end{align}
which is satisfied by the results in \cref{eq: s c_s and c_V}. In the above, the entropy density $s$ is always positive, thus $c_s^2$ and $c_V$ should have an equal sign. Since thermodynamical stability requires that $c_V>0$, so hydrodynamical stability needs $c_s^2>0$, which equals to $p+q\leq 4$. Thus the allowed values for $p$ and $q$ are $p=1$, $q=1,2,3$; $p=2$, $q=1,2$ and $p=3$, $q=1$. The total number of the allowed cases is 6, which are in one-to-one correspondence with the cases of compactified Dp-brane \cite{Wu2012}. We will explain this by the end of the next section.

\section{The second-order results}

The results of the second-order constraint relations and the Navier-Stokes equations can be found in the appendix.

To solve the second-order perturbations, we need to expand the global on-shell metric \cref{eq: global on-shell metric} to the second order as
\begin{align}\label{eq: 2nd order expanded metric}
  ds^2 =& -r^\frac{9-p-q}{p} \bigg[ f - (1-f) \delta\beta_i \delta\beta_i - \frac{(7-p-q)r_H^{6-p-q}}{r^{7-p-q}} (\delta r_H + \frac12 \delta^2r_H + \delta r_H^{(1)}) \cr
  & - \frac{(7-p-q)(6-p-q) r_H^{5-p-q}}{2r^{7-p-q}}(\delta r_H)^2 + (F_k+\delta F_k) \partial \beta + F_k(\delta\partial \beta + \delta\beta_i \partial_0\beta_i) \cr
  & + 2 a(r)\delta\beta_i \partial_0\beta_i + k^{(2)}(r) \bigg] dv^2 + 2r^\frac{9-p-q}{p} \bigg[ (f-1)(\delta\beta_i + \frac12 \delta^2\beta_i) \cr
  & + a(\partial_0\beta_i + \delta\partial_0\beta_i + \delta\beta_j\partial_j\beta_i) - \frac{(7-p-q)r_H^{6-p-q}}{r^{7-p-q}}\delta r_H \delta\beta_i + F_k \partial \beta \delta\beta_i \cr
  & - F \delta\beta_j \partial_{(i}\beta_{j)} + w_i^{(2)}(r) \bigg] dvdx^i + 2r^\frac{(p-3)(p-6) + q(p-2)}{2p} \bigg[ 1 + (F_j+\delta F_j) \partial \beta \cr
  &  + F_j(\delta\partial \beta + \delta\beta_i\partial_0\beta_i) + \frac12 \delta\beta_i\delta\beta_i + j^{(2)}(r) \bigg] dvdr \cr
  & + r^\frac{9-p-q}{p} \bigg[ \delta_{ij} + (1-f)\delta\beta_i\delta\beta_j - 2a \delta\beta_{(i}\partial_{|0|}\beta_{j)} + (F + \delta F) \partial_{(i}\beta_{j)} \cr
  & + F \left( \delta\partial_{(i} \beta_{j)} + \delta\beta_{(i} \partial_{|0|}\beta_{j)} \right) + \alpha_{ij}^{(2)}(r) + h^{(2)}(r) \delta_{ij} \bigg] dx^idx^j \cr
  & - 2 r^\frac{(p-3)(p-6) + q(p-2)}{2p} \bigg( \delta\beta_i + \frac12 \delta^2\beta_i + F_j \partial\beta \delta\beta_i \bigg) dx^i dr,
\end{align}
where we have defined
\begin{align}
  \delta \mathcal F(r_H(x),r) = - \frac{(5-p-q) \mathcal{F}(r) + 2r \mathcal{F}'(r)}{2r_H} \delta r_H,
\end{align}
with $\mathcal F$ referring to any of the $F,~F_j$, or $F_k$. The second-order metric perturbations can be solved by putting \cref{eq: 2nd order expanded metric} into the Einstein equations. The solving procedure is similar to our previous works \cite{Wu1604,Wu1608,Wu1807,Wu2012,Wu2111} and will be omitted here.

The second-order constitutive relations of the relativistic fluids dual to the smeared Dp-brane can be written as
\begin{align}
  T_{\mu\nu} =&\; \frac{1}{2 \kappa_{p+2}^2} \Bigg\{ {r_H^{7-p-q} \over L_p^{8-p-q}} \left( \frac{9-p-q}{2} u_\mu u_\nu + \frac{5-p-q}{2} P_{\mu\nu} \right) \cr
  & - \left( \frac{r_H}{L_p} \right) ^\frac{9-p-q}{2} \bigg( 2\sigma_{\mu\nu} + \frac{2(p-3)^2 + 2q(p-1)}{p(9-p-q)} P_{\mu\nu} \partial u \bigg) \cr
  & + \frac{r_H^2}{L_p} \Bigg[ \bigg( \frac{1}{5-p-q} + \frac{1}{7-p-q} H_\frac{5-p-q}{7-p-q} \bigg)\cdot 2\bigg( \sideset{_\langle}{}{\mathop D}\sigma_{\mu\nu\rangle} + \frac1{p} \sigma_{\mu\nu} \partial u \bigg) \cr
  & + \bigg( \frac{3(p-3)^2 + 3q(p-1)}{(5-p-q)(9-p-q)} - \frac{(p-3)^2 + q(p-1)}{(7-p-q)(9-p-q)} H_\frac{5-p-q}{7-p-q} \bigg) \frac{2\sigma_{\mu\nu} \partial u}{p} \cr
  & + \frac{1}{5-p-q} \cdot 4\sigma_{\langle\mu}^{~~\rho}\sigma_{\nu\rangle\rho} + \bigg( - \frac{2}{5-p-q} + \frac{2}{7-p-q} H_\frac{5-p-q}{7-p-q} \bigg) \cdot 2 \sigma_{\langle\mu}^{~~\rho} \Omega_{\nu\rangle\rho} \Bigg] \cr
  & + \frac{r_H^2}{L_p} P_{\mu\nu} \Bigg[ \bigg( \frac{2(p-3)^2 + 2q(p-1)}{p(5-p-q)(9-p-q)} + \frac{2(p-3)^2 + q(p-1)}{p(7-p-q)(9-p-q)} H_\frac{5-p-q}{7-p-q} \bigg) D(\partial u) \cr
  & + \Bigg( \frac{[2(p-3)^2 + 2q(p-1)] [(3p^2 - 17p + 18) + q(3p-2)]}{p^2 (5-p-q)(9-p-q)^2}  \cr
  & + \frac{(5-p-q) [2(p-3)^2 + 2q(p-1)]}{p (7-p-q) (9-p-q)^2} H_\frac{5-p-q}{7-p-q} \Bigg) (\partial u)^2 \cr
  & + \frac{(p-3)^2 + q(p-1)}{p(5-p-q)(9-p-q)} \cdot 4 \sigma_{\alpha\beta}^2 \Bigg] \Bigg\}.
\end{align}
Then all the dynamical second-order transport coefficients can be read as
\begin{align}\label{eq: 2nd order transport coefs}
  \eta\tau_\pi &= \frac1{2\kappa_{p+2}^2} \left( \frac{1}{5-p-q} + \frac{1}{7-p-q} H_\frac{5-p-q}{7-p-q} \right) \frac{r_H^2}{L_p}, \cr
  \eta\tau_\pi^* &= \frac1{2\kappa_{p+2}^2} \left[ \frac{3(p-3)^2 + 3q(p-1)}{(5-p-q)(9-p-q)} - \frac{(p-3)^2 + q(p-1)}{(7-p-q)(9-p-q)} H_\frac{5-p-q}{7-p-q} \right] \frac{r_H^2}{L_p}, \cr
  \lambda_1 &= \frac1{2\kappa_{p+2}^2} \frac{1}{5-p-q} \frac{r_H^2}{L_p}, \qquad \lambda_2 = \frac1{2\kappa_{p+2}^2} \left( - \frac{2}{5-p-q} + \frac{2}{7-p-q} H_\frac{5-p-q}{7-p-q} \right) \frac{r_H^2}{L_p}, \cr
  \zeta\tau_\Pi &= \frac1{2\kappa_{p+2}^2} \Bigg[ \frac{2(p-3)^2 + 2q(p-1)}{p(5-p-q)(9-p-q)} + \frac{2(p-3)^2 + q(p-1)}{p(7-p-q)(9-p-q)} H_\frac{5-p-q}{7-p-q} \Bigg] \frac{r_H^2}{L_p}, \cr
  \xi_1 &= \frac1{2\kappa_{p+2}^2} \frac{(p-3)^2 + q(p-1)}{p(5-p-q)(9-p-q)} \frac{r_H^2}{L_p}, \cr
  \xi_2 &= \frac1{2\kappa_{p+2}^2} \Bigg[ \frac{[2(p-3)^2 + 2q(p-1)] [(3p^2 - 17p + 18) + q(3p-2)]}{p^2 (5-p-q)(9-p-q)^2} \cr
  & \quad + \frac{(5-p-q) [2(p-3)^2 + 2q(p-1)]}{p (7-p-q) (9-p-q)^2} H_\frac{5-p-q}{7-p-q} \Bigg] \frac{r_H^2}{L_p}.
\end{align}

From the first- and second-order transport coefficients of the smeared brane, we can see that substituting $p$ for $p-q$ will bring the results here back to the form of compactified Dp-brane \cite{Wu1604,Wu2012}. So we find an interesting correspondence between the compactified Dp-brane and the smeared Dp-brane, that is, the results of D(p+q)-brane are in one-to-one correspondence with D[(p+q)-q]-brane, i.e.
\begin{align}
  \text{Results of D(p+q)-brane} \quad \longleftrightarrow \quad \text{Results of D[(p+q)-q]-brane}.
\end{align}
Please note that both the D(p+q)-brane and the D[(p+q)-q]-brane are dual to $(1+p)$-dimensional relativistic fluid. Let's take $p=3$ and $q=1$ as an example, it is the D3-brane uniformly smeared on 1 transverse direction, i.e. the D(3+1)-brane. Its results are completely the same as D(4-1)-brane, i.e., the D4-brane with 1 direction compactified. In this way, the 6 cases of the smeared Dp-brane are in one-to-one correspondence with the compactified Dp-brane.

The reason for the correspondence between the smeared and compactified Dp-brane is that these two kinds of branes are actually connected by T-dual. A T-dual on a compact transverse direction of one Dp-brane makes a D(p+1)-brane with one world-volume direction compact, i.e. the D[(p+1)-1]-brane in our notation. Then a T-dual on a compact transverse direction with many Dp-branes uniformly distributed will give us the same amount of D[(p+1)-1]-branes. So the T-dual on $q$ compact transverse directions which have Dp-branes uniformly distributed shows the equivalence between the D(p+q)-brane and the D[(p+q)-q]-brane.

\section{Discussions and outlook}

In this paper, we investigate the Dp-brane uniformly distributed on $q$ compact transverse directions. After integrating the $(8-p-q)$-dimensional unit sphere and the $q$-dimensional smeared dimensions, the smeared Dp-brane background becomes a $(p+2)$-dimensional gravity coupled with one scalar field. This $(p+2)$-dimensional reduced theory finally turns out to be a Chamblin-Reall model. Thus, we have found 4 Chamblin-Reall models which can give exact and analytic results to the second-order transport coefficients of non-conformal relativistic fluids. They are the reduced compactified AdS black hole \cite{Wu2111}, the Dp-brane \cite{Wu1807}, the compactified Dp-brane \cite{Wu2012}, and the smeared Dp-brane.

We also calculate all the 7 dynamical second-order transport coefficients of the smeared Dp-brane. We find the results are in one-to-one correspondence with the compactified Dp-brane, so the smeared Dp-brane is equal to the compactified Dp-brane. The reason for this equivalence is that these two backgrounds are actually connected by T-dual. So up to now, the number of non-conformal gravity backgrounds which can be exactly solved should be 3.

After continuous searches on the non-conformal and exactly solvable gravity backgrounds, we have finally found 3 kinds of such backgrounds. This will clearly establish the direction for our future research works. For example, if we would like to know the 4 non-trivial thermal second-order transport coefficients for the solvable non-conformal backgrounds. We just need to do calculations for the reduced AdS black hole, the Dp-brane and the compactified Dp-brane.

This paper also discusses the CSC on smeared Dp-brane from a new thermal-hydro viewpoint. This is different from previous studies on the same topic \cite{Bostock0405,Ross0503,Harmark0509}, which all relate the local thermal stability with the classical stability of metric perturbations. The hydrodynamical stability gives a constraint for the value of $p$, $q$ and the allowed cases are 6, which is in one-to-one correspondence with the previously studied compactified Dp-brane.

\section*{Acknowledgement}

C. Wu would like to thank Dr. Qiang Jia for very helpful discussions. We also thank the Young Scientists Fund of the National Natural Science Foundation of China (Grant No. 11805002) for support.

\appendix

\section{The reduction ansatz}

The reduction ansatz we use for the smeared Dp-brane is
\begin{align}
  ds^2 = e^{2\alpha_1 A} g_{MN} dx^M dx^N + e^{2\alpha_2 A}\left( e^{2\beta_1 B} \delta_{mn}dy^m dy^n + e^{2\beta_2 B} L_p^2 d\Omega_{8-p-q}^2 \right).
\end{align}
The $y^m$ are the coordinates of the transverse directions to the branes on which they are delocalized. The Christoffel symbols can be calculated as
\begin{align}
  \widetilde\Gamma^M_{NP} &= \Gamma^M_{NP} + \alpha_1 ( \delta^M_N\partial_P A + \delta^M_P\partial_N A - g_{NP}\nabla^M A ), \cr
  \widetilde\Gamma^M_{mn} &= - (\alpha_2\nabla^M A + \beta_1\nabla^M B) e^{(-2\alpha_1 + 2\alpha_2)A + 2\beta_1B}\delta_{mn}, \cr
  \widetilde\Gamma^n_{Mm} &= (\alpha_2\partial_M A + \beta_1\partial_M B) \delta_m^n, \cr
  \widetilde\Gamma^M_{ab} &= - (\alpha_2\nabla^M A + \beta_2\nabla^M B) e^{(-2\alpha_1 + 2\alpha_2)A + 2\beta_2B} \gamma_{ab} L_p^2, \cr
  \widetilde\Gamma^a_{Mb} &= (\alpha_2\partial_M A + \beta_2\partial_M B) \delta^a_b, \cr
  \widetilde\Gamma^a_{bc} &= {}^{\Omega}\Gamma^a_{bc}.
\end{align}
Here the $\widetilde \Gamma^{\hat M}_{\hat N \hat P}$ and $\Gamma^M_{NP}$ are separately the Christoffel symbols in 10 and $p+2$ dimensions. The ${}^{\Omega}\Gamma^a_{bc}$ are the Christoffel symbols of the $(8-p-q)$-dimensional unit sphere. The following relations will be very useful in calculating the Ricci tensor
\begin{align}
  \widetilde\Gamma^{N}_{MN} &= \Gamma^N_{MN} + (p+2) \alpha_1\partial_M A, \cr
  \widetilde\Gamma^{\hat N}_{M\hat N} &= \Gamma^N_{MN} + [(p+2)\alpha_1 + (8-p)\alpha_2] \partial_M A + [q \beta_1 + (8-p-q)\beta_2] \partial_M B.
\end{align}

Then the Ricci tensor can be got as the following
\begin{align}
  \mathcal R_{MN} =&\; R_{MN} - [p\alpha_1 + (8-p)\alpha_2] \nabla_M \nabla_N A - \alpha_1 g_{MN} \nabla_P \nabla^P A  \cr
  &  - [q \beta_1 + (8-p-q)\beta_2] \nabla_M\nabla_N B \cr
  &+ \left[ p \alpha_1^2 + 2(8-p) \alpha_1\alpha_2 - (8-p) \alpha_2^2 \right] \partial_M A\partial_N A \cr
  & - \left[ p \alpha_1^2 + (8-p) \alpha_1\alpha_2 \right] g_{MN} (\partial A)^2 \cr
  &+ (\alpha_1 - \alpha_2) [ q \beta_1 + (8-p-q) \beta_2 ] (\partial_M A \partial_N B + \partial_N A \partial_M B) \cr
  &- \alpha_1 [ q \beta_1 + (8-p-q) \beta_2 ] g_{MN} \partial_P A \partial^P B \cr
  &- \left[q \beta_1^2 + (8-p-q) \beta_2^2 \right] \partial_M B \partial_N B; \\
  \mathcal R_{mn} =& - \left[ \alpha_2 \nabla^2 A + \beta_1 \nabla^2 B + \left( p \alpha_1 \alpha_2 + (8-p) \alpha_2^2 \right) (\partial A)^2 \right. \cr
  & + \left. \left( p \alpha_1 \beta_1 + (8-p+q) \alpha_2 \beta_1 + (8-p-q) \alpha_2 \beta_2 \right) \partial A \partial B \right. \cr
  & \left.+ \left( q \beta_1^2 + (8-p-q) \beta_1\beta_2 \right) (\partial B)^2 \right] e^{(-2\alpha_1 + 2\alpha_2) A + 2 \beta_1 B} \delta_{mn}; \\
  \mathcal R_{ab} =&\ (7-p-q) \gamma_{ab} - \left[ \alpha_2\nabla^2 A + \beta_2 \nabla^2 B + \left( p \alpha_1 \alpha_2 + (8-p) \alpha_2^2\right) (\partial A)^2  \right. \cr
  &\left. + \left( p \alpha_1\beta_2 + q \alpha_2\beta_1 + (16-2p-q) \alpha_2 \beta_2 \right) \partial A \partial B \right. \cr
  &\left. + \left( q \beta_1 \beta_2 + (8-p-q) \beta_2^2\right) (\partial B)^2 \right] e^{(- 2 \alpha_1 + 2 \alpha_2) A + 2 \beta_2 B} \gamma_{ab} L_p^2.
\end{align}
Here $\mathcal R_{\hat M \hat N}$ and $R_{MN}$ are separately the 10- and $(p+2)$-dimensional Ricci tensor. Finally the Ricci scalar can be calculated as
\begin{align}
   \mathcal R =&\; e^{-2\alpha_1 A} \Big[ R - 2 ((p+1) \alpha_1 + (8-p) \alpha_2) \nabla^2 A - 2 (q \beta_1 + (8-p-q) \beta_2) \nabla^2 B \cr
  & - \big( p(p+1) \alpha_1^2 + 2p(8-p) \alpha_1\alpha_2 + (8-p)(9-p) \alpha_2^2 \big) (\partial A)^2 \cr
  & - 2 \big( pq\, \alpha_1\beta_1 + p(8-p-q) \alpha_1\beta_2 + q(9-p) \alpha_2\beta_1 + (8-p-q)(9-p) \alpha_2\beta_2 \big) \partial A \partial B \cr
  & - \big( q(q+1) \beta_1^2 + 2q(8-p-q) \beta_1 \beta_2 + (8-p-q)(9-p-q) \beta_2^2 \big) (\partial B)^2 \Big] \cr
  & + \frac{(7-p-q)(8-p-q)}{L_p^2} e^{-2\alpha_2 A - 2\beta_2 B}.
\end{align}

\section{A universal treatment of the first-order tensor perturbations for the Chamblin-Reall models with one background scalar}

In solving the first-order tensor perturbation, we have set $\alpha_{ij}^{(1)} (r) = F(r) \sigma_{ij}$. The differential equations of $F(r)$ depend on $p$ in the cases of Dp-brane \cite{Wu1807}, compactified Dp-brane \cite{Wu1604,Wu2012} and compactified AdS black hole \cite{Wu2111}. For the smeared Dp-brane, $F$ also contains $q$. In our previous works on Dp- \cite{Wu1807} and compactified Dp-brane \cite{Wu2012}, we have solved $F$ for every allowed $p$ case by case. But we only offer the situations for $2\leq p \leq 5$ in \cite{Wu2111}, cases of $p\geq 6$ are not covered. Since the expressions of $F$ play a key role in solving the perturbations: all the first-order scalar perturbations can be expressed through $F$. The search for a general expression of $F$ valid for all the models with all the allowed values of $p$ is necessary. Luckily we find a universal treatment for the solution of $F$.

The differential equations for $F(r)$ can be written in the form like
\begin{align}
  \frac{d}{dr} \left( r^{\alpha+1} f(r) \frac{dF}{dr} \right) = - \gamma r^\beta, \qquad f(r) = 1 - \frac{r_H^\alpha}{r^\alpha}.
\end{align}
The solution can be written in the form that
\begin{align}
  F(r) =& \int_{\infty}^{r} \frac{dx}{x^{\alpha+1} f(x)} \int_{r_H}^{x} (-\gamma y^\beta) dy \cr
    =& - \frac{\gamma}{\beta+1} \left[ \int_{\infty}^{r} x^{\beta-\alpha} \left( 1 - \frac{r_H^\alpha}{x^\alpha} \right)^{-1} dx - r_H^{\beta+1} \int_{\infty}^{r} \frac{dx}{x^{\alpha+1} f(x)} \right].
\end{align}
The second integral in the above bracket is easy to get:
\begin{align}
  \int_{\infty}^{r} \frac{dx}{x^{\alpha+1} f(x)} = \frac{1}{\alpha r_H^\alpha} \ln f(r).
\end{align}
In order to do the first one, we need to use the substitution $x = r t^\frac{1}{\alpha}$, such that
\begin{align}
  \int_{\infty}^{r} x^{\beta-\alpha} \left( 1 - \frac{r_H^\alpha}{x^\alpha} \right)^{-1} dx = - \frac{1}{\alpha r^{\alpha-\beta-1}} \int_{1}^{\infty} t^{\frac{\beta+1}{\alpha} - 1} \left( t - \frac{r_H^\alpha}{r^\alpha} \right)^{-1} dt.
\end{align}
Now with the help of the integral expression for the hypergeometric function
\begin{align}
  B(a,c-a) \; {}_2F_1\left( a,b;c,x \right) = \int_{1}^{\infty} t^{b-c} (t-1)^{c-a-1} (t-x)^{-b} dt,
\end{align}
one has
\begin{align}
  \int_{\infty}^{r} x^{\beta-\alpha} \left( 1 - \frac{r_H^\alpha}{x^\alpha} \right)^{-1} dx = - \frac{1}{(\alpha-\beta-1) r^{\alpha-\beta-1}} {}_2F_1\left( \frac{\alpha-\beta-1}{\alpha}, 1; \frac{\alpha-\beta-1}{\alpha}+1, \frac{r_H^\alpha}{r^\alpha} \right).
\end{align}
Here $B(a,b) = \int_0^1 t^{a-1} (1-t)^{b-1} dt = \Gamma(a) \Gamma(b) / \Gamma(a+b)$ is the beta function. Thus $F(r)$ can be finally read as
\begin{align}\label{eq: general solution of F}
  F(r) =&\; \frac{\gamma}{(\beta+1) (\alpha-\beta-1) r^{\alpha-\beta-1}} \, {}_2F_1\left( 1, \frac{\alpha-\beta-1}{\alpha}; 1+\frac{\alpha-\beta-1}{\alpha}, \frac{r_H^\alpha}{r^\alpha} \right) \cr
  & + \frac{\gamma \ln f(r)}{\alpha (\beta+1) r_H^{\alpha-\beta-1}}.
\end{align}
\cref{tab: parameters in all the CR backgrounds} gives the explicit values of $\alpha$, $\beta$, and $\gamma$ for all the first-order tensor perturbations in the Chamblin-Reall backgrounds with one scalar field.
\begin{table}[h!]
\centering
\begin{tabular}{|c|c|c|c|}
  \hline
                                                &  $\alpha$  &  $\beta$               &  $\gamma$   \\ \hline
  Dp-brane                               &  $7-p$      &  $\frac{7-p}{2}$  &  $9-p$  \\ \hline
  compactified Dp-brane          & $7-p$      &   $\frac{7-p}{2}$  &  $9-p$  \\ \hline
  smeared Dp-brane                 & $7-p-q$   & $\frac{7-p-q}{2}$ &  $9-p-q$  \\ \hline
  compactified AdS black hole &  $p+1$    &           $p-1$          &  $2p$ \\
  \hline
\end{tabular}
\caption{\label{tab: parameters in all the CR backgrounds} The value of $\alpha$, $\beta$, and $\gamma$ for the differential equations of the first-order tensor perturbations for all the Chamblin-Reall backgrounds with one scalar field.}
\end{table}

Next, we need to prove that \cref{eq: general solution of F} satisfied the following two boundary conditions:
\begin{itemize}
\item[(1)] $F(r \to \infty) \to 0$
\item[(2)] $F(r_H)$ is finite
\end{itemize}
The first one is easy to prove. As $r \to \infty$, $r_H^\alpha / r^\alpha \to 0$. Then $\ln f(r) \to 0$ and ${}_2F_1 \to 1$, since one has ${}_2F_1\left( a,b; c, 0 \right) = 1$. One should also note that $\alpha-\beta-1$ is always larger than 0 for the cases in \cref{tab: parameters in all the CR backgrounds}, thus we have $F(r \to \infty) \to 0$.

To prove the second one, we have to figure out the expansion of ${}_2F_1\left( 1, a; 1+a, \frac{1}{x} \right)$ around 1. To the lowest order, the result is
\begin{align}
  {}_2F_1\left( 1, a; 1+a, \frac{1}{x} \right) \bigg|_{x=1+\epsilon} \approx -a (\gamma_E + \psi(a) + \ln \epsilon),
\end{align}
where $\epsilon$ is a positive infinitesimal, $\gamma_E$ is the Euler constant, and $\psi(x) = \frac{d \ln \Gamma(x)}{dx}$ is the digamma function. Now setting $a = \frac{\alpha-\beta-1}{\alpha}$ and $\frac{1}{x} = \frac{r_H^\alpha}{r^\alpha}$, we can expand $F$ around $x=1$ as
\begin{align}
  F(r)\bigg|_{x=1+\epsilon} \approx &\; \frac{\gamma}{(\beta+1) (\alpha-\beta-1) r_H^{\alpha-\beta-1}} \left[ - \frac{\alpha-\beta-1}{\alpha} \left( \gamma_E + \psi\left( \frac{\alpha-\beta-1}{\alpha} \right) + \ln \epsilon \right) \right] \cr
  & + \frac{\gamma}{\alpha (\beta+1) r_H^{\alpha-\beta-1}} \ln \epsilon \cr
  =& - \frac{\gamma}{\alpha(\beta+1) r_H^{\alpha-\beta-1}} \left( \gamma_E + \psi\left( \frac{\alpha-\beta-1}{\alpha} \right) + \ln \epsilon \right) + \frac{\gamma}{\alpha(\beta+1) r_H^{\alpha-\beta-1}} \ln \epsilon \cr
  =& - \frac{\gamma}{\alpha(\beta+1) r_H^{\alpha-\beta-1}} \left( \gamma_E + \psi\left( \frac{\alpha-\beta-1}{\alpha} \right)  \right)
\end{align}
One can see that the infinite part in the expansion of the hypergeometric function exactly cancels the term in $\ln f(r)$, leaving us a finite result.

Thus \cref{eq: general solution of F} is the correct expression of $F$ for all the Chamblin-Reall models with one background scalar. Specifically for the compactified AdS black hole \cite{Wu2111}, the solution \cref{eq: general solution of F} covers all the cases of $p \geq 2$ and $1 \leq q \leq p-1$.

\section{The second-order constraint relations and the Navier-Stokes equation}

In deriving the second-order differential equations and the Navier-Stokes equations, one needs the second-order constraint relations derived from $\partial_\mu \partial^\rho T^{(0)}_{\rho\nu} = 0$:
\begin{align}
  & \frac{9-p-q}{2} \frac{1}{r_H} \partial _0^2 r_H + \partial_0\partial \beta - \frac{2}{9-p-q} (\partial \beta)^2 - \frac{4}{5-p-q} \partial_0\beta_i \partial_0\beta_i = 0,  \\
  & \frac{5-p-q}{2} \frac{1}{r_H} \partial_i^2 r_H + \partial_0 \partial \beta - \frac{2}{5-p-q} \partial_0\beta_i \partial_0\beta_i - \frac{5-p-q}{9-p-q} (\partial \beta)^2 + \partial_i \beta_j \partial _j\beta_i = 0,  \\
  & \frac{5-p-q}{2} \frac{1}{r_H} \partial _0\partial _i r_H + \partial _0^2\beta_i - \frac{7-p-q}{9-p-q} \partial _0\beta_i \partial \beta + \partial_0 \beta_j \partial _j\beta_i = 0, \\
  & \frac{9-p-q}{2} \frac{1}{r_H} \partial _0\partial _i r_H + \partial_i \partial \beta - \frac{2}{5-p-q} \partial _0\beta_i \partial \beta - \frac{4}{5-p-q} \partial_0 \beta_j \partial _i\beta_j = 0, \\
  & \partial _0 \Omega_{ij} - \frac{5-p-q}{9-p-q} \Omega_{ij} \partial \beta - \partial_k\beta_{[i} \partial_{j]}\beta_k = 0, \\
  & \frac{5-p-q}{2} \frac{1}{r_H} \partial_i \partial _j r_H + \partial _0 \partial_{(i}\beta_{j)} - \frac{2}{5-p-q} \partial_0\beta_i \partial_0\beta_j - \frac{5-p-q}{9-p-q} \partial_{(i}\beta_{j)} \partial\beta + \partial _k\beta _{(i} \partial _{j)} \beta _k = 0
\end{align}
Then using the notation defined in \cref{tab: 2nd order spatial viscous terms},
\begin{table}[h]
\centering
\begin{tabular}{|l|l|l|}
  \hline
  Scalars of $\mathrm{SO}(p)$                                 &                 Vectors of $\mathrm{SO}(p)$                      &\quad  Tensors of $\mathrm{SO}(p)$ \\ \hline\hline
 $\mathbf{s}_1=\frac1{r_H}\partial_0^2r_H$                  & $\mathbf{v}_{1i} = \frac1{r_H} \partial_0\partial_i r_H$ & $\mathbf{t}_{1ij} = \frac1{r_H} \partial_i\partial_j r_H - \frac1{p} \delta_{ij} \mathbf{s}_3$ \\
  $\mathbf{s}_2 = \partial_0\partial_i\beta_i$                 & $\mathbf{v}_{2i} = \partial_0^2\beta_i$       & $\mathbf{t}_{2ij} = \partial_0 \Omega_{ij}$ \\
  $\mathbf{s}_3 = \frac1{r_H}\partial_i^2r_H$             & $\mathbf{v}_{3i} = \partial_j^2\beta_i$        & $\mathbf{t}_{3ij} = \partial_0\sigma_{ij}$ \\
  $\mathfrak S_1 = \partial_0\beta_i\partial_0\beta_i$     & $\mathbf{v}_{4i}=\partial_j\Omega_{ij}$      & $\mathfrak T_{1ij} = \partial_0\beta_i\partial_0\beta_j - \frac1{p} \delta_{ij} \mathfrak S_1$ \\
  $\mathfrak S_2 = \epsilon_{ijk} \partial_0 \beta_i \partial_j \beta_k$       & $\mathbf{v}_{5i} = \partial_j\sigma_{ij}$      & $\mathfrak T_{2ij} = \sigma_{[i}^{~~k} \Omega_{j]k}$ \\
  $\mathfrak S_3 = (\partial_i\beta_i)^2$                        & $\mathfrak V_{1i} = \partial_0\beta_i\partial\beta$     & $\mathfrak T_{3ij} = \Omega_{ij} \partial \beta$ \\
  $\mathfrak S_4 = \Omega_{ij} \Omega_{ij}$              &  $\mathfrak V_{2i} = \partial_0\beta_j \Omega_{ij}$  & $\mathfrak T_{4ij}=\sigma_{ij}\partial\beta$ \\
  $\mathfrak S_5=\sigma_{ij}\sigma_{ij}$                      & $\mathfrak V_{3i} = \partial_0\beta_j \sigma_{ij}$  & $\mathfrak T_{5ij} = \Omega_i^{~k}\Omega_{jk} - \frac1{p} \delta_{ij} \mathfrak S_4$ \\
                                                                                    &                                                             & $\mathfrak T_{6ij} = \sigma_i^{~k}\sigma_{jk} - \frac1{p} \delta_{ij} \mathfrak S_5$ \\
                                                                                    &                                                             & $\mathfrak T_{7ij} = \sigma_{(i}^{~~k} \Omega_{j)k}$ \\
\hline
\end{tabular}
\caption{\label{tab: 2nd order spatial viscous terms} All the $\mathrm{SO}(p)$ invariant second-order spatial viscous terms of the relativistic fluid dual to the smeared Dp-brane for the cases of $1\leq p \leq 3$. We do not have $\mathfrak T_{2,5,6}$ when $p=2$ and we only have $\mathbf s _{1,2,3},~\mathfrak S_{1,3}, ~\mathbf v_{1,2,3}$ and $\mathfrak V_1$ at $p=1$.}
\end{table}
we can reexpress the above constraint relations by
\begin{align}
  & \mathbf s_1 + \frac2{9-p-q} \mathbf s_2 - \frac{8}{(9-p-q)(5-p-q)} \mathfrak S_1 - \frac{4}{(9-p-q)^2} \mathfrak S_3 = 0,  \\
  & \mathbf s_2 + \frac{5-p-q}{2} \mathbf s_3 - \frac{2}{5-p-q} \mathfrak S_1 + \frac{(p-3)^2+q(p-1)}{p(9-p-q)} \mathfrak S_3 - \mathfrak
  S_4 + \mathfrak S_5 = 0,  \\
  & \mathbf v_1 + \frac{2}{5-p-q} \mathbf v_2 + \frac{2(p^2-8p+9 + q(p-1))}{p(9-p-q)(5-p-q)} \mathfrak V_1 - \frac{2}{5-p-q} \mathfrak V_2 + \frac{2}{5-p-q} \mathfrak V_3 = 0,  \\
  & \mathbf v_1 + \frac{2p (\mathbf v_4 + \mathbf v_5)}{(p-1)(9-p-q)} - \frac{4(p+2) \mathfrak V_1}{p(9-p-q)(5-p-q)} - \frac{8 (\mathfrak V_2 + \mathfrak V_3)}{(9-p-q)(5-p-q)} = 0, \\
  & \mathbf t_2 - 2 \mathfrak T_2 + \frac{p^2-7p+18 + q(p-2)}{p(9-p-q)} \mathfrak T_3 = 0,  \\
  & \mathbf t_1 + \frac{2}{5-p-q} \mathbf t_3 - \frac{4}{(5-p-q)^2} \mathfrak T_1 + \frac{2(p^2-7p+18) + 2q(p-2)}{p(9-p-q)(5-p-q)} \mathfrak T_4  \cr
  & ~~\, - \frac{2}{5-p-q} \mathfrak T_5 + \frac{2}{5-p-q} \mathfrak T_6 = 0.
\end{align}

One also gets the Navier-Stokes equation through the conservation equation $\partial^\mu T^{(0+1)}_{\mu\nu} = 0$:
\begin{align}
  \frac{1}{r_H^{(p+q-3)/2}} \partial_0 r_H^{(1)} =&\; \frac{4(p-3)^2 + 4q(p-1)}{p(9-p-q)^2(7-p-q)} \mathfrak S_3 + \frac{4}{(9-p-q)(7-p-q)} \mathfrak S_5, \\
  \frac{1}{r_H^{(p+q-3)/2}} \partial_i r_H^{(1)} =&\; \frac{[4(p-3)^2 + 4q(p-1)]  \mathbf v_4 + 16p  \mathbf v_5}{(p-1)(9-p-q)(7-p-q)(5-p-q)} \cr
 & + \frac{2(p+q-1)((p+q)^2 - 22(p+q) + 77) - 2q((p+q)^2 - 22(p+q) + 85)}{p(9-p-q)(7-p-q)(5-p-q)^2} \mathfrak V_1 \cr
 & - \frac{2(19 - 3p-3q)}{(9-p-q)(7-p-q)(5-p-q)} \mathfrak V_2 \cr
 & - \frac{2(p^2 - 14p + 77)-2q(14-2p-q)}{(9-p-q)(7-p-q)(5-p-q)^2} \mathfrak V_3.
\end{align}

\section{The calculation on the smeared D0-brane}

Ref. \cite{Ross0503} has proved that the D0-branes smeared on $p$ transverse directions are both thermally and dynamically unstable by the method in \cite{Reall0104}. But as we have mentioned that ref. \cite{Emparan1205} points out that thermodynamical stability should relate to hydrodynamical stability. So here we review this point for smeared D0-brane through the fluid/gravity correspondence.

The 10-dimensional action for the smeared D0-brane is
\begin{align}
  S =&\; \frac1{2\kappa_{10}^2} \int d^{10}x \sqrt{-G} \left[ \mathcal R - \frac12 (\nabla_{\hat M}\phi)^2 - \frac{g_s^2}{2\cdot 8!} e^{- \frac32 \phi} \tilde F_{\hat M_1 \cdots \hat M_8}^2 \right].
\end{align}
It is the on-shell action of the following 10-dimensional background
\begin{align}
   ds^2 &= - H^{-\frac{7}{8}} f(r) dt^2 + H^\frac18 \left( d \vv x^2 + \frac{dr^2}{f(r)} + r^2 d\Omega_{8-p}^2 \right), \cr
   e^\phi &= H^{\frac34}, \qquad \qquad \tilde F_{x^1 \cdots x^p \theta^1 \cdots \theta^{8-p}}= g_s^{-1} Q \sqrt{\gamma_{8-p}}.
\end{align}
Here $H=1+\left(\frac{r_0}{r}\right)^{7-p}$, and the parameters are related by $Q=(7-p)L^{7-p}$ and $L^{2(7-p)} = r_0^{7-p} (r_0^{7-p} + r_H^{7-p})$. Under the extremal limit, the background becomes
\begin{align}
   ds^2 &= - \left(\frac{r}{L}\right)^\frac{7(7-p)}{8} f(r) dt^2 + \left(\frac{L}{r}\right)^\frac{7-p}{8} \left( d\vv x^2 + \frac{dr^2}{f(r)} + r^2 d\Omega_{8-p}^2 \right), \cr
   e^\phi &=  \left( \frac rL \right)^{- \frac{3(7-p)}{4}}, \qquad \qquad \tilde F_{x^1 \cdots x^p \theta^1 \cdots \theta^{8-p}}= g_s^{-1} Q \sqrt{\gamma_{8-p}}.
\end{align}

With the reduction ansatz
\begin{align}\label{eq:  smeared D0 reduction ansatz}
  ds^2 = e^{2\alpha_1 A} g_{MN} dx^M dx^N + L^2 e^{2\alpha_2 A} d\Omega_{8-p}^2,
\end{align}
the parameters in the above are $\alpha_1 = -\frac{8-p}{p}$ and $\alpha_2 = 1$. We can get the reduced theory as
\begin{align}
  S &= \frac1{2\kappa_{p+2}^2} \int d^{p+2}x \sqrt{-g} \left[ R - \frac12 (\partial \phi)^2 - \frac{8(8-p)}{p} (\partial A)^2 - V(\phi,A) \right], \cr
  V &= \frac{(7-p)^2}{2L^2} e^{\frac{9+p}{6(7-p)} \phi - \frac{2(8-p)}{p} A} -  \frac{(7-p)(8-p)}{L^2} e^{-\frac{16}{p} A}.
\end{align}
and
\begin{align}\label{eq: smeared D0 reduced background}
   ds^2 &= - \left(\frac rL \right)^{\frac{9+6p-p^2}{p}} f(r) dt^2 + \left(\frac rL \right)^\frac{9-p}{p} \left( d\vv x^2 + \frac{dr^2}{f(r)} \right) \cr
   e^\phi &= \left( \frac rL \right)^{- \frac{3(7-p)}{4}}, \qquad \qquad e^A = \left(\frac rL\right)^\frac{9+p}{16}
\end{align}
The result of the Ricci scalar of \cref{eq:  smeared D0 reduction ansatz} can be borrowed from \cite{Wu1807}. From the above one can see that $\phi$ and $A$ are not independent of each other. We can replace $A$ with $\phi$ via $A = - \frac{9+p}{12(7-p)} \phi$, then the reduced action becomes
\begin{align}
  S = \frac{1}{2\kappa_{p+2}^2} \int d^{p+2}x \sqrt{-g} \left[ R - \frac{4(p+1)(9-p)^2}{9p(7-p)^2}(\partial \phi)^2 + \frac{(7-p)(9-p)}{2L^2} e^{\frac{4(9+p)}{3p(7-p)} \phi} \right].
\end{align}
One can still recast the above into the Chamblin-Reall form by redefining the scalar field as
\begin{align}
  \sqrt{\frac{4(p+1)(9-p)^2}{9p(7-p)^2}} \phi = \frac{1}{\sqrt2} \varphi
\end{align}
Thus the Chamblin-Reall form of the reduced action reads
\begin{align}
  S = \frac{1}{2\kappa_{p+2}^2} \int d^{p+2}x \sqrt{-g} \left[ R - \frac12 (\partial \varphi)^2 + \frac{(7-p)(9-p)}{2L^2} e^{\gamma \varphi} \right],
\end{align}
with the background now is
\begin{align}
  ds^2 &= - \left(\frac rL \right)^{\frac{9+6p-p^2}{p}} f(r) dt^2 + \left(\frac rL \right)^\frac{9-p}{p} \left( d\vv x^2 + \frac{dr^2}{f(r)} \right) \cr
   e^\varphi &= \left( \frac rL \right)^{- \frac{9+p}{p \gamma}}.
\end{align}
Here $\gamma^2 = \frac{2(9+p)^2}{p(p+1)(9-p)^2}$. In the above metric of the reduced background, the factor in front of $- f(r) dt^2$ is different from that in front of $d \vv x^2$, this means there is no Lorentz invariance in the directions of $(t, \vv x)$, thus this reduced metric does not have dual fluid. So from the viewpoint of the fluid/gravity correspondence, the smeared D0-brane does not dual to any relativistic fluid.


\providecommand{\href}[2]{#2}\begingroup\raggedright\endgroup

\end{document}